\documentclass[traditabstract]{aa} 
\usepackage{epsfig,amsmath}
\usepackage[dvips]{color}
\usepackage{verbatim}
\usepackage{graphicx}
\usepackage{txfonts}
\usepackage{natbib}
\bibpunct{(}{)}{;}{a}{}{,}
\usepackage{hyperref}
\usepackage{longtable}
\usepackage{lscape}
\newcommand{\Mpc}{\textrm{Mpc}}

\newcommand{\km}{\textrm{km}}
\newcommand{\cm}{\textrm{cm}}
\begin{document}
   \title{A Ly$\alpha$ blob and $z_{\rm abs}\approx z_{\rm em}$ damped Ly$\alpha$ absorber in the dark matter halo of the binary quasar Q\,0151+048\thanks{Based on observations done with {\it i)} European Southern Observatory (ESO) utilizing 8.2m Very Large Telescope (VLT) X-shooter spectrograph on Cerro Paranal in the Atacama Desert, northern Chile. {\it ii)} 2.56m Nordic Optical Telescope
(NOT), a scientific association between Denmark, Finland, Iceland,
Norway and Sweden, operated at Observatorio del Roque de Los
Muchachos on the island of La Palma, Spain.}}

\author{Tayyaba Zafar\inst{1}
 \and Palle M\o ller \inst{2}
  \and C\'{e}dric Ledoux \inst{3}
 \and Johan P.U. Fynbo\inst{1,2}
    \and Kim K. Nilsson \inst{4}
   \and Lise Christensen\inst{2}
   \and Sandro D'Odorico \inst{2}
    \and Bo Milvang-Jensen \inst{1}
    \and Micha{\l} J. Micha{\l}owski\inst{1,5}
    \and Desiree D. M. Ferreira \inst{1} } 

\institute{Dark Cosmology Centre, Niels Bohr Institute, University of Copenhagen,
Juliane Maries Vej 30, DK-2100 Copenhagen \O, Denmark.
\and European Southern Observatory, Karl-Schwarzschildstrasse 2  D-85748 Garching, Germany
 \and European Southern Observatory, Alonso de C\'{o}rdova 3107, Vitacura, Casilla 19001 Santiago 19, Chile
 \and ST-ECF, Karl-Schwarzschildstrasse 2, 85748, Garching bei Munchen, Germany
 \and Scottish Universities Physics Alliance, Institute for Astronomy,
University of Edinburgh, Royal Observatory, Edinburgh, EH9 3HJ, United Kingdom} 
\authorrunning{Tayyaba Zafar et al.}
\titlerunning{Q\,0151+048 system at $z=1.93$}
\offprints{tayyaba@dark-cosmology.dk}

\date{Received  / Accepted }

\abstract
{Q\,0151+048 is a physical quasar (QSO) pair at $z\sim1.929$ with a
separation of 3.3 arcsec on the sky. In the spectrum of the brighter
member of this pair, Q\,0151+048A, a damped Ly$\alpha$ absorber
(DLA) is observed at a higher redshift. We
have previously detected the host galaxies of both QSOs as well as a
Ly$\alpha$ blob whose emission surrounding Q\,0151+048A extends over
$5\times 3.3$ arcsec.
We seek to constrain the geometry of the system and understand the
possible relations between the DLA, the Ly$\alpha$ blob and the two
QSOs. We also aim at characterizing the former two objects in more
detail.
To study the nature of the Ly$\alpha$ blob, we have performed
low-resolution long-slit spectroscopy with the slit aligned with the
extended emission. We have also observed the whole system using the
medium-resolution VLT/X-shooter spectrograph and the slit aligned with
the two QSOs. The systemic redshift of both QSOs is determined from
rest-frame optical emission lines redshifted into the NIR. We employ
line-profile fitting techniques, to measure metallicities and the
velocity width of low-ionization metal absorption lines associated to
the DLA, and photo-ionization modeling, to characterize the DLA
further.
We measure systemic redshifts of $z_{em(\rm{A})}=1.92924\pm0.00036$
and $z_{em(\rm{B})}=1.92863\pm0.00042$ from the $\ion{H}{\beta}$ and
$\ion{H}{\alpha}$ emission lines, respectively. In other words, the
two QSOs have identical redshifts within $2\sigma$. From the width of
Balmer emission lines and the strength of the rest-frame optical
continuum, we estimate the masses of the black holes of the two QSOs
to be 10$^{9.33}$ M$_{\odot}$ and 10$^{8.38}$ M$_{\odot}$ for
Q\,0151+048A and Q\,0151+048B, respectively. We then use the
correlation between black hole mass and dark matter halo mass to infer
the mass of the dark matter halos hosting the two QSOs: 10$^{13.74}$
M$_{\odot}$ and 10$^{13.13}$ M$_{\odot}$ for Q\,0151+048A and
Q\,0151+048B, respectively. We observe a velocity gradient along
the major axis of the Ly$\alpha$ blob consistent with the rotation
curve of a large disk galaxy, but it may also be caused by gas inflow
or outflow. We detect residual continuum in the DLA trough which we
interpret as emission from the host galaxy of Q\,0151+048A. The
derived ${\rm H}^0$ column density of the DLA is $\log
N_{{\rm H}^0}=20.34\pm0.02$ $\cm^{-2}$. Metal column densities are
also determined for a number of low-ionization species resulting in an
overall metallicity of 0.01 Z$_{\odot}$. We detect $\ion{C}{ii}^\ast$ which allows 
us to make a physical model of the DLA cloud.
From the systemic redshifts of the QSOs, we conclude that the
Ly$\alpha$ blob is associated with Q\,0151+048A rather than with the
DLA. The DLA must be located in front of both the Ly$\alpha$ blob and
Q\,0151+048A at a distance larger than 30 kpc and has a velocity relative to the blob of $640 \pm 70$ km/s. The two quasars accrete at normal Eddington ratios. The DM halo of
this double quasar will grow to the mass of our local super-cluster
at $z=0$. We point out that those objects therefore form an ideal
laboratory to study the physical interactions in a $z=2$ pre-cursor of
our local super-cluster.}
\maketitle{}
\section{Introduction}

Searches for Ly$\alpha$ emission in the high redshift Universe have in the 
last decade found many sources of a type called Ly$\alpha$ ``blobs'' 
\citep{fynbo,keel,steidel,francis,matsuda,palunas,weidinger04,dey,villar,nilsson}.
 Ly$\alpha$ blobs must be considered
the extreme end of the distribution of Ly$\alpha$ emitting sources, in that they
are very extended (sizes more than several times ten kpc across), and unusually bright
in the Ly$\alpha$ line ($\log{Ly\alpha} \gtrsim 43$~erg~s$^{-1}$). They are 
interesting and promising cosmological probes, partly because they are some of the largest
cohesive structures at high redshift and partly because they
hold key insights into the energetics of their source. Among
the Ly$\alpha$ blobs, a number of sub-categories have emerged, sorting
them based on the presumed main source of ionizing flux. Several blobs
have been shown to be powered by vigorous star formation \citep{taniguchi,ohyama,mori,matsuda2006}), others by cold gas falling onto a dark matter (DM) halo \citep{nilsson,smith}.

In this group of extreme objects, the most extreme sub-sample in terms of sizes and 
fluxes are those powered by active galactic nuclei (AGN), QSOs and radio galaxies. Gigantic Ly$\alpha$ haloes 
are routinely discovered around high redshift radio galaxies (see \citet{miley} for 
a review). Smaller blobs are also often powered by AGN activity; \citet{geach} find 
that roughly $17$\% of narrow-band selected Ly$\alpha$ blobs are powered by AGN.  In especially
rare cases, multiple component systems including a QSO, a DLA,
and a Ly$\alpha$ blob are found \citep{moller,leibundgut,hennawi}. It is yet unclear 
how these systems interact, or even their actual spatial configurations. 

One such system is the binary QSO Q\,0151+048A \& B (also called
PHL1222), whose brightest member was first observed spectroscopically by \citet{burbidge} who measured
its redshift to be $z=1.943$ (at that time the other member of the binary was
not yet discovered). \citet{williams} were the first to discuss the nature of
the $z_{abs} > z_{em}$ DLA in the spectrum of
PHL1222 in some detail. They inferred a redshift of the absorber of more than
2000 km s$^{-1}$ relative to the redshift of Q\,0151+048A (hereafter qA).
The DLA is hence a member of the class of $z_{abs} \approx z_{em}$ DLAs
\citep[see][]{moller} which later became proximate DLAs (PDLAs)
(\citet{ellison}; see also
\citet{weymann,prochaska2008}). Compared to normal intervening DLAs PDLAs tend to have
evidence for a higher radiation field (presumably the proximate QSO) and
on average slightly higher metallicities \citep{ellison}.

The QSO redshift reported in the above work was measured from the
UV emission lines from
${\rm N}^{4+}$, ${\rm Si}^{3+}$ and ${\rm C}^{3+}$, which are now known to be
systematically blueshifted relative to the systemic redshift \citep{tytler}.
\citet{williams} also inferred a minimum distance of 0.4 Mpc between the DLA
and qA based on the absence of fine structure lines from ${\rm Si}^+$ and
${\rm C}^+$. Later \citet{meylan} were the first to establish that PHL1222 has
a fainter companion at nearly the same redshift. The pair was discussed in more
detail in \citet{moller} where Ly$\alpha$ emission was also detected in the DLA
trough. \citet[][hereafter FMW99]{fynbo} found using narrow band imaging that
there is very extended Ly$\alpha$ emission close to the redshift of the DLA and in
retrospect this appears to be one of the first detections of what is now known
as Ly$\alpha$ blobs.

In this study we present new spectroscopic observations of both members of the
QSO pair. Our main objectives are {\it i)} to characterize the DLA absorber (in
particular its metallicity and velocity profile), {\it ii)} to determine the
systemic redshift of both members of the pair using rest frame optical emission
lines, {\it iii)} further constrain the geometry of the system by looking for
absorption in the spectrum of Q\,0151+048B (qB) at the redshift of the DLA, and
{\it iv)} to further study the nature of the Ly$\alpha$ emission blob.

The outline of the paper is as follows: In $\S$2 we describe our
spectroscopic observations and data reduction. In $\S$3 \& 4 we analyze
emission features (the Ly$\alpha$ emitting gas and QSO emission lines)
and absorption signatures of the DLA respectively and in \S5 we
discuss how these new data change our interpretation of the geometry and
dynamical state of the system. In $\S$6 we briefly summarize our conclusions.
Throughout this paper, we assume a Hubble constant of H$_0=72$ $\km$ s$^{-1}
~\Mpc^{-1}$ \citep{freedman} and a universe with flat cosmology $\Omega_m=0.27$,
$\Omega_\Lambda=0.73$ following the results presented in the initial WMAP data release \citep{spergel}. 
At a redshift of 1.93 and with this cosmology, 1 arcsec
corresponds to a proper length of 8.384 kpc. All wavelengths are corrected to
vacuum heliocentric frame.

\section{Observations and data reduction}

\subsection{The NOT/ALFOSC data}

\begin{figure}
	\centering 
	{\includegraphics[width=8.5cm,clip=]{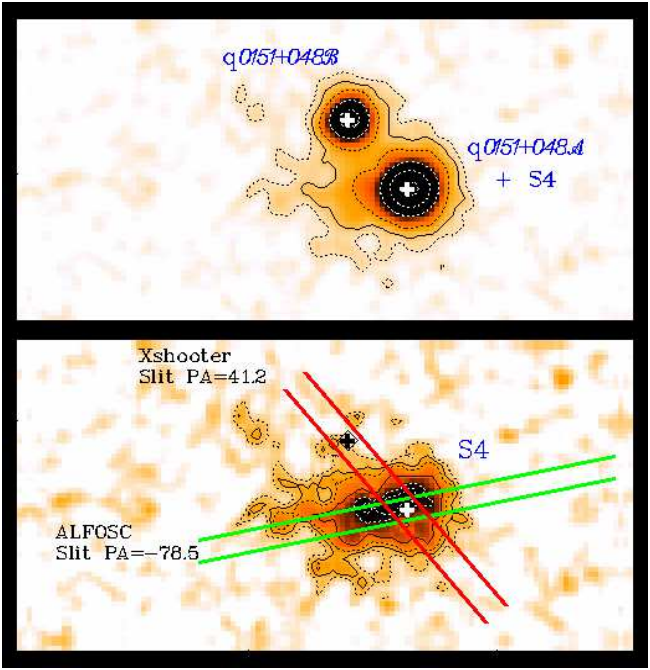} }
	\caption{Top frame is a cut-out of the original narrow
band image (FMW99). Contours are shown in black and the centres of
the two QSOs are marked by a white cross. NE is to the upper left.
In the bottom frame the QSOs were psf-subtracted and only the extended
Ly$\alpha$ emission is seen. The crosses still mark the position
of the QSOs. The slit-position along the Ly$\alpha$ major axis
used for the NOT/ALFOSC observations is shown in green, the
X-shooter A+B slit-position is shown in red.} 
		\label{slitpos} 
\end{figure}

On three nights in September 1997 we used the NOT equipped with Andalucia Faint Object Spectrograph and Camera (ALFOSC) to obtain a spectrum of
the elongated Ly$\alpha$ emission line object discovered by
FMW99 (see their Fig.~2) who named it S4. Our slit was centered on qA and
aligned with the major axis of S4 at a position angle of $-78.5$ deg
(see Fig.~\ref{slitpos}). We used ALFOSC grism G6 and a slit of $0.7''$
which provided a resolution of $R=\lambda/\Delta \lambda=700$. We integrated for a
total exposure time of 33000 sec at this slit position.
A detailed log is provided in Table~\ref{log}.
The data were reduced using standard techniques for bias and flat-fielding
and a final stack was obtained using optimal weights.

\begin{table}
\begin{minipage}[t]{\columnwidth}
\setlength{\tabcolsep}{4.5pt}
\caption{Log of spectroscopic Observations of Q\,0151+048 A \& B with NOT (ALFOSC) and VLT/X-Shooter (XSH).}      
\label{log} 
\centering     
\renewcommand{\footnoterule}{}  
\begin{tabular}{@{}c c c c c c c@{} }  
\hline\hline                        
Date & Object & Exp. & Instrument & slit & R & PA \\
 &  & (sec)  & &width & & (deg)\\
\hline
2 Sep, 97 & A+S4 & 14000 & ALFOSC & $0.7''$  & 700 & $-78.5$\\
3 Sep, 97 & A+S4 & 15000 & ALFOSC & $0.7''$  & 700 & $-78.5$\\
4 Sep, 97 & A+S4 &  4000 & ALFOSC & $0.7''$  & 700 & $-78.5$\\
\hline
18 Nov, 08 & A+B & 3600 & XSH/UVB\footnote{UVB-arm range: 3000-5500\,\AA{}} & $0.8''$  & 6200 & $41.2$\\
   & A+B & 3600 & XSH/VIS\footnote{VIS-arm range: 5500-10000\,\AA{}} & $0.7''$ & 11000 & 41.2 \\
\hline
29 Sept, 09 & A+B & 960 & XSH/UVB &$1.0''$ & 5100 & 41.2 \\
   & A+B & 960 & XSH/VIS & $0.9''$  & 8800 & 41.2\\
   & A+B & 480 & XSH/NIR & $0.9''$  & 5100 & 41.2 \\
\hline
\end{tabular}
\end{minipage}
\end{table}

\subsection{VLT/X-shooter data}

On November 18th 2008, during the first instrument commissioning run,
we returned to this field, this time using the X-shooter on
the VLT which allowed us to observe both QSOs with higher
spectral resolution and long wavelength coverage from ultraviolet (UV) to near-infrared (NIR).
The ESO-VLT X-shooter is an Echelle spectrograph mounted at the
VLT Cassegrain focus \citep{odorico}. Here we used a slit PA of
41.2 deg which placed both QSOs on the slit
(see Fig.~\ref{slitpos}) thereby allowing us to obtain a 3600 seconds
exposure of both QSOs (Program ID: 60.A-9022(C)).
During this early commissioning run only two of the three arms
were operational and we obtained spectra in parallel with the
UV-blue (UVB) and visual (VIS) arms of the spectrograph using slits of $0.8''$
and $0.7''$ respectively. The resulting resolutions are 6200 (UVB)
and 11000 (VIS).
Because the seeing was significantly larger than the slit widths
the sky-line widths provide a good measure of the resolution.

On September 29, 2009, during a Science Verification run
(Program ID: 60.A-9441(A)), we again integrated with the
X-shooter slits on both QSOs.
This time we could utilize all three arms (i.e. UVB, VIS, and
NIR) and the log of observations is again provided in
Table~\ref{log}. Availability of the NIR arm during
the September run allowed us to extend the wavelength coverage through
to rest wavelength optical lines. We obtained 480 sec in the
NIR arm with a resolution of $R=5100$. All X-shooter data
are available from the ESO archive.

The X-shooter data were processed using a preliminary version of the X-shooter data reduction pipeline \citep{goldoni,modigliani}. The pipeline first corrected UVB and VIS raw frames for bias and NIR frames for dark current. After division by the flat field, the pipeline performed background subtraction, corrected cosmic ray hits using the method developed by \citet{van} and subtracted sky emission lines using the \citet{kelson} method. The individual orders were extracted and rectified in wavelength space. This was done by using the previously obtained wavelength solution from calibration frames. The individual orders were merged afterwards and in the overlapping regions the merging was weighted by the errors which were being propagated during the reduction process. From the final 2D merged spectrum one dimensional spectra for qA \& qB were extracted.

We obtained a spectrum of the hot B-type main sequence star
Hip020971 with the NIR arm on the same night. The spectrum was
extracted using the same procedure and was subsequently used
for atmospheric absorption correction.
No flux calibration was applied to either of the spectra.


\section{Results}

\subsection{The NOT/ALFOSC data}

\begin{figure}
	\centering 
	{\includegraphics[angle=-90,width=\columnwidth,clip=]{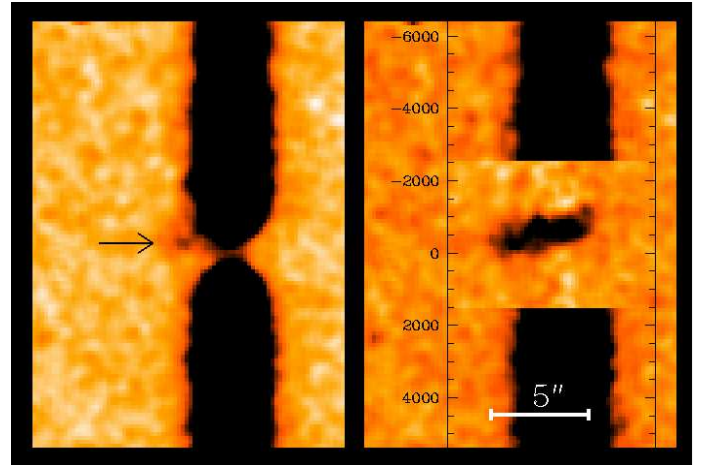} }
	\caption{{\it Left panel:} NOT/ALFOSC 2D spectrum of qA,
the dispersion is vertical with short wavelengths at the top.
The DLA line is at the centre, the arrow points to faint
extended Ly$\alpha$ emission.
{ \it Right panel:} Same as left after SPSF subtraction. The full
extent of the Ly$\alpha$ emission is now visible. It is seen to be tilted
and shifted relative to the DLA. The vertical axis provides relative
velocity (km/s) in the rest frame of the DLA absorber.} 
		\label{notpsf} 
\end{figure}

The purpose of the NOT spectrum was to study the dynamics of the
Ly$\alpha$ emitting gas (S4). The resolution of the spectrum is too
low to provide any useful information about the DLA metallicity, so
the primary interest is in the region around Ly$\alpha$ emission.
In Fig.~\ref{notpsf} (left panel) we show this region of the 2D stacked, sky
subtracted spectrum. We here use a negative representation where
black is bright and orange is zero. In the centre the DLA absorption trough
is a dominant feature, and the arrow points to a small but significant
area of extended line emission. To examine the Ly$\alpha$ emission
closer to the QSO it is necessary first to remove the emission
from the QSO, a technique referred to as Spectral-PSF (or SPSF)
subtraction \citep{mollerpsf, moller2000}.

\begin{figure}
	\centering 
	{\includegraphics[width=\columnwidth,bb=18 190 595 676,clip=]{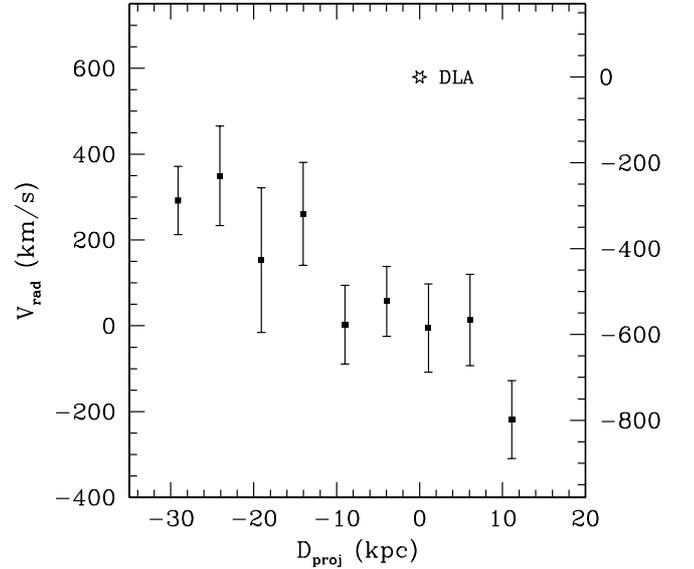} }
	\caption{Velocity field of the Ly$\alpha$ emission seen in the right panel of Fig.~\ref{notpsf}. The DLA absorption sightline towards the qA is marked by a $``\star"$. } 
		\label{velocity} 
\end{figure}

In the right panel of Fig.~\ref{notpsf} we show the same part of the spectrum
after SPSF subtraction around the extended Ly$\alpha$ emission.
We clearly detect extended Ly$\alpha$ over a total region
of about 5 arcsec (42 kpc proper), corresponding well to the extend of the emission
seen in narrowband imaging (Fig.~\ref{slitpos}, lower panel). It is also clear that the
emission is not symmetrically distributed around the QSO (also
consistent with Fig.~\ref{slitpos}), that it is well separated (in redshift)
from the DLA and, somewhat surprisingly, that the emission has a
strong velocity gradient along the major axis which has a remarkable
resemblance to a rotation curve. The velocity gradient as measured on
the SPSF subtracted spectrum is shown in Fig.~\ref{velocity}.

\begin{figure}
	\centering 
	{\includegraphics[width=\columnwidth,clip=]{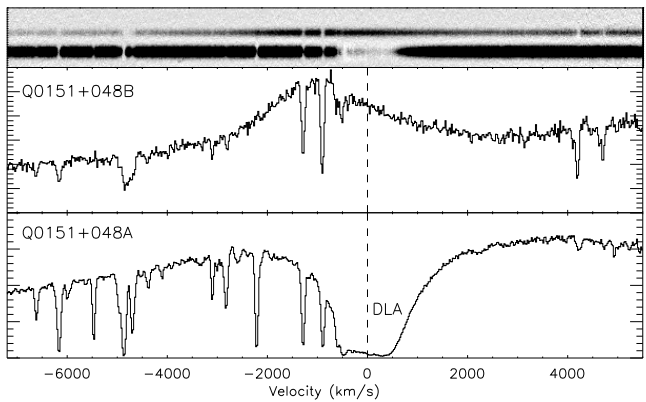} }
	\caption{X-shooter spectra of both QSOs obtained on November, 18 2008
with a single exposure of 3600 sec.
{\it Upper panel:} 2D UVB-arm spectra of both QSOs, qB is the fainter object and
qA is the brighter object displaying strong DLA absorption.
{ \it Middle panel:} 1D spectrum of qB.
{ \it Lower panel:} 1D spectrum of qA showing the DLA absorption. Note that in both the 2D and the
1D spectrum of qA residual flux is clearly visible in the damped trough.}
		\label{dla} 
\end{figure} 

\subsection{VLT/X-shooter data}
\subsubsection{Residual flux in the DLA trough}
One of the most striking features of the X-shooter spectrum
of qA is that the DLA absorption line does not go completely
to zero. Ly$\alpha$ emission in or close to DLA absorption lines
has been seen in several cases. \citet{mw}, \citet{pettini},
\citet{djorgovski}, \citet{leibundgut}, \citet{moller2002},
\citet{moller2004} and \citet{fynbo10} all reported a narrow
emission line slightly offset from the position of the background
quasar and consistent with being Ly$\alpha$ emission from the DLA
host which absorb the QSO to zero flux at Ly$\alpha$.
\citet{cooke} report what
they interpret as either a broad absorption line with emission at the
centre or as two absorption lines separated by non-absorbed
quasar continuum. Unfortunately they do not provide the detailed
spatial information which would help the interpretation, but
since a possible interpretation of the ``feature'' is that it is
non-absorbed quasar continuum it must be centered on the position
of the quasar and have the spatial profile of a point-source. This
makes it unlikely that it is Ly$\alpha$ emission as Ly$\alpha$
emission usually is more extended than a point-source and in
addition it would be an unlikely coincidence that the Ly$\alpha$
emission by chance would be lined up exactly with the second
image of the lens where the DLA covers only the first image.

\citet{hennawi} also report an emission line in a damped
absorption line. They do not provide any information about spatial
offset, but the object has a large (5 arcsec) size along the slit.
\citet{hennawi} rule out that the emission is related to the DLA and
instead interpret it as a Ly$\alpha$ blob physically connected to the
quasar. This object has two features in common with the DLA in qA.
Both are PDLAs and both have Ly$\alpha$ emission extending 5 arcsec.
The extended Ly$\alpha$ of qA is however significantly offset
in redshift from the PDLA and does not enter into the absorption line. Nevertheless
in both the 2D and the 1D X-shooter spectrum of qA (Fig.~\ref{dla})
it is clearly seen that there is a smooth continuum providing a
non-zero ``floor'' in the Ly$\alpha$ line. In this section we aim
to determine the source of this residual flux in the damped
absorption line.

There are two possible ways to explain this residual flux. Either
the absorber is so small that it only partly covers the emitter
(the central engine), or there are at least two different sources
of emission and the DLA absorber is only covering one of those.
In \citet{fynbo2} it was reported that qA is located
inside a comparatively bright host galaxy with a $B$($\rm{AB}$) magnitude
of $20.8\pm0.2$ and $I$($\rm{AB}$) of $21.5\pm0.2$. The host was best fitted
by a de Vaucouleurs profile, and in $B$($\rm{AB}$) it was found to be 3.0
magnitude fainter than qA while in $u$($\rm{AB}$) and $I$($\rm{AB}$) it was 3.9 and
4.0 magnitude fainter respectively. We now investigate if
the residual flux in the DLA trough simply is the spectrum of the
QSO host galaxy.

We first co-added the profile of all wavelength bins in the centre
of the DLA line where it should be completely absorbed. We then did
the same where the QSO continuum had recovered both on the blue
side ($3540-3545$\,\AA) and on the red side ($3595-3605$\,\AA). The
FWHM of the QSO profile along the slit on both the blue and the
red side is 1.03 arcsec while the FWHM of the flux at the bottom
of the DLA is 1.57 arcsec, i.e. significantly wider.
Simple subtraction in squares gives an intrinsic FWHM of 1.2 arcsec
for the object seen in the DLA line which
immediately rules out that the residual could be caused by partial
coverage of the QSO alone because in this case the profile should
be the same as that of the QSO. The object that we see in the
DLA trough is therefore a different object than the QSO, it is
relatively large (1.2 arcsec FWHM) and could possibly be the host galaxy.

The spectrum was taken with a slit of $0.8''$  and we therefore
have slit losses of both the QSO and of the other object. We have
modeled the psf of the QSO such that the psf-on-slit is identical
to that observed. Using this we measure a slit loss of $40\%$. Similarly
we have modeled the profile of the object seen in the trough, and find
slit losses of $60\%$ in this case. The measured counts per bin of the
extended object is a factor 27 less than of the QSO continuum.
Correcting for the different slit losses this becomes a factor 18
fainter, i.e. 3.1 magnitudes fainter. This corresponds well to the
reported magnitude difference as found via imaging \citep{fynbo2}.
From Fig.~\ref{dla} upper panel (more clearly seen in
Fig.~\ref{twodspec}) it is also seen that the two objects are centered
on the same position.

Summing up all of the evidence, extended profile, centered on the QSO,
similar brightness relative to qA, we conclude that the flux seen in
the DLA trough is the signature of the object reported to be the host
in \citep{fynbo2}. It remains to be determined whether the flux from
the host is stellar emission or dust reflection. The gradient of the
emission in the trough could suggest that it at least partly is quasar
light reflected off dust in the host halo. Spectro-polarimetric data 
would be required to resolve this question.

\begin{table*}
\caption{Emission line properties of Q\,0151+048A \& B. The columns provide observed and restframe wavelengths, observed and restframe equivalent widths with their corresponding errors, line IDs and emission redshifts.}            
\label{emission}    		
\centering                                  
\begin{tabular}{c c c c c c c c } 	
\hline\hline
Object  & $\lambda_{obs}$ & $\lambda_{rest}$ & $EW_{obs}$ & $\sigma_     {EW}$ & $EW_{rest}$& ID & $z_{em}$ \\
 &  $\AA$ & $\AA$ & $\AA$ & $\AA$ & $\AA$& &  \\
\hline
qA: & 3612.1710 & 1240.13 & 13.658    & 0.237  & 4.688 & $\ion{N}{v}$ & $1.91274\pm0.00014$  \\
    & 3694.1012 & 1262.59 & 3.9000 & 0.090 &  1.330  & $\ion{Si}{ii}$ &  $1.92581\pm0.00027$ \\
    & 3814.8126 & 1304.35 & 11.400  &  0.113  & 3.900 & $\ion{O}{i}$ & $1.92468\pm0.00035$ \\
    & 3894.1912 & 1334.50 & 3.8000 & 0.087 & 1.300 & $\ion{C}{ii}$  & $1.91809\pm0.00016$  \\
    & 4063.6774 & 1399.70 &  40.350  & 0.198  & 13.90 & $\ion{Si}{iv} + \ion{O}{iv}$ & $1.90325\pm0.00030$ \\
    & 4504.6833 & 1549.06 &  62.700  & 0.266   & 21.56 & $\ion{C}{iv}$ &  $1.90801\pm0.00040$  \\
    & 6804.3376 & UV 344(2325.38) & 19.160 & 0.110 & 6.550 & \ion{Fe}{ii} & $1.92612\pm0.00026$ \\
    & 7109.3696 & UV 360(2433.436) & 10.730 & 0.140 & 3.670 & \ion{Fe}{ii} & $1.92154\pm0.00038$  \\
    & 8188.7607 & 2798.75 & 96.440 & 0.240 & 32.96 & $\ion{Mg}{ii}$& $1.92586\pm0.00040$   \\
    & 14244.000 & 4862.68 & 119.80 & 0.280 & 40.90 & $\ion{H}{\beta}$ & $1.92924\pm0.00036$  \\
    & 19231.400 & 6564.61 & 525.56 & 0.810 & 179.4 & $\ion{H}{\alpha}$ & $1.92939\pm0.00043$  \\
\hline 
qB: & 3557.3960 & 1215.67 &  243.70  &  0.923 & 83.300 & Ly$\alpha$ & $1.92628 \pm 0.00034$\\
    & 3628.3042 & 1240.14 &   8.2900  &  0.631 & 2.8300 &  $\ion{N}{v}$ & $1.92572 \pm 0.00042$\\
    & 3678.1892 & 1262.59 &  2.1300   &  0.409  & 0.7300 & $\ion{Si}{ii}$ &  $1.91321\pm0.00045$\\
    & 3810.1611 & 1304.35 &  2.3840  &  1.009  & 0.8190 & $\ion{O}{i}$ & $1.92112\pm 0.00036$ \\
    & 4094.7356 & 1399.70 &  32.023  &  0.963 & 10.986 & $\ion{Si}{iv} + \ion{O}{iv}$ & $1.92544\pm0.00025$ \\
    & 4528.4208 & 1549.06 &  60.699  &  1.039  & 20.752 & $\ion{C}{iv}$ &  $1.92333\pm0.00043$ \\
    & 8198.4067 & 2798.75 & 89.800 & 0.270 & 30.650 & $\ion{Mg}{ii}$& $1.92931\pm0.00039$  \\
    & 19225.300 & 6564.61 & 492.72 & 2.580 & 168.20 & $\ion{H}{\alpha}$ & $1.92863\pm0.00042$ \\
\hline
\end{tabular}
\end{table*} 

\subsubsection{Emission line analysis of Q\,0151+048A \& B}
   \begin{figure}
   \centering 
   {\includegraphics[width=\columnwidth,clip=]{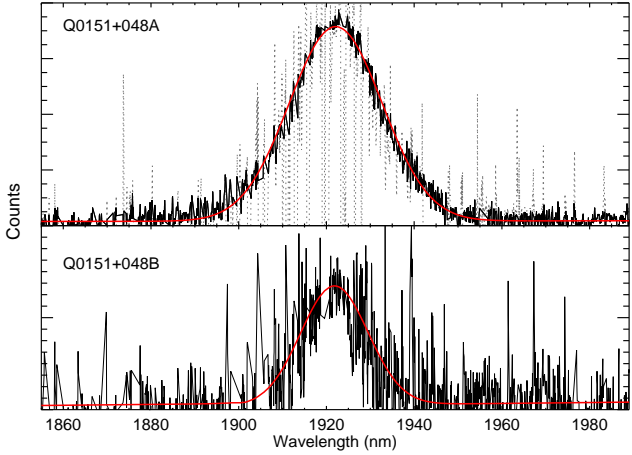}}
      \caption{$\ion{H}{\alpha}$ emission lines from Q\,0151+048 A \& B after subtraction of the QSOs continuum.
As described in the text only the good data intervals between the strong
airglow residuals were used. The data, including airglow residuals are
shown as a light dotted line, the solid black line is data with the
not used parts removed.
The solid red lines show the best gaussian profile fit to the used
intervals. The inferred redshifts are $z_{\ion{H}{\alpha}(A)}=1.9294$ and $z_{\ion{H}{\alpha}(B)}=1.9286$.}
          \label{Fignir}
   \end{figure}

  \begin{figure}
   \centering
   {\includegraphics[width=\columnwidth,clip=]{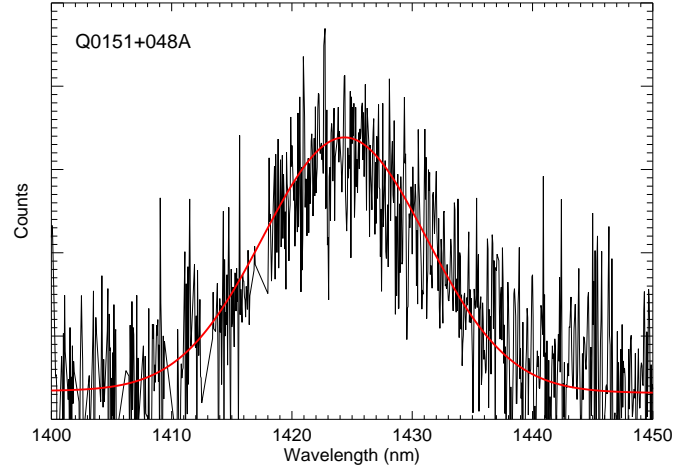}}
      \caption{$\ion{H}{\beta}$ emission line from Q\,0151+048 A after the continuum subtraction. The solid red line shows the best gaussian fit (see caption to Fig.~\ref{Fignir} for details) providing an emission redshift of $z_{\ion{H}{\beta}(A)}=1.9292$.}
          \label{Figbeta}
   \end{figure}

The original interpretation of the Ly$\alpha$ emission from this
system was severely complicated by the large apparent blueshift of qA
relative to the DLA (1250 km s$^{-1}$) as well as to the
Ly$\alpha$ emitter. Such shifts are common in QSOs
spectra \citep[see][]{grandi, wilkes, mw, laor, mcintosh, vanden,
richards}, but even though there is a
general pattern to the shifts it has not yet been possible to
device a method to successfully correct for this. We therefore
had serious doubts about the accuracy of QSO systemic
redshifts determined from UV emission lines, and one of the two
main drivers for the X-shooter spectroscopy was to obtain systemic
redshifts based on rest-frame optical emission lines. This is possible because
of the X-shooter NIR arm. In Fig.~\ref{Fignir} and \ref{Figbeta} we
show the $\ion{H}{\alpha}$ and $\ion{H}{\beta}$ emission lines respectively. For qB, 
the fainter of the two QSOs, only $\ion{H}{\alpha}$ was detected. The
blue part of the $\ion{H}{\alpha}$ emission lines was affected by
atmospheric absorption which corrected out well with the use of
hot B-type main sequence stars.

In Table~\ref{emission} we list observed and rest-frame parameters
for the emission lines of the spectra of qA \& qB. The continuum is well detected for both QSOs and the equivalent widths are measured by integrating the emission line between the end points and above the estimated continuum. For the
$\lambda_{obs}$ and inferred $z_{em}$ we use the same peak (or
mode) definition as in \citet{moller}. It is seen that the inferred redshifts of qA range from
1.9034 to 1.92939 corresponding to a relative velocity
of 2700 km s$^{-1}$ vindicating our doubts about the previous redshift
determination.

The narrow and unblended $[\ion{O}{iii}]\,\lambda 5007$
emission line is presumed to trace the systemic (center of mass)
redshift of the galaxy \citep{gaskell, vrtilek, mcintosh}.
Unfortunately the $[\ion{O}{iii}]$ emission lines at our
redshift are in a region of strong telluric absorption. The
telluric absorption combined with the very narrow profile of
$[\ion{O}{iii}]$ makes it impossible to recover the lines
via telluric correction. The $\ion{H}{\alpha}$ line is in a
part of the spectrum which is even stronger absorbed, but
the width of the line is here helpful in recovering parts
of the line. The $\ion{H}{\beta}$ line is intrinsically
fainter than the $\ion{H}{\alpha}$ but in a region of less
telluric absorption and therefore easier to recover.

In addition to the telluric absorption there are many strong
sky emission lines present causing large residuals after
sky subtraction. With the high resolution of X-shooter and for
the wide $\ion{H}{\alpha}$ and $\ion{H}{\beta}$ lines this is not
problematic as we simply masked out the affected intervals and
used only the remaining sections (see Fig.~\ref{Fignir}).
We subtracted continuum from the emission lines and 
then proceeded to fit a smooth profile to the central part
of the lines using only the intervals not masked out. After
the fit to the data we used the $1\sigma$ error spectrum to create
1,000 random realizations of errors,
added those to the spectrum and repeated the process. In Table~\ref{emission} we list the $standard~deviation$
of this distribution as statistical errors on the redshift.
For qA we compute the optimally combined redshift as the
weighted mean of the two lines and obtain $z_{em(A)}=1.92930\pm 0.00028$,
for qB we only have $\ion{H}{\alpha}$ and therefore simply
adopt the redshift of that line as the systemic
$z_{em(B)}=1.92863\pm 0.00042$. Redshifts determined from those
lines are known to have an offset of only a few tens of km s$^{-1}$
from the true systemic redshift of the QSO \citep[][]{vanden}.

\subsubsection{Extended emission in the X-shooter spectrum}

\begin{figure}
	\centering
	{\includegraphics[width=\columnwidth,clip=]{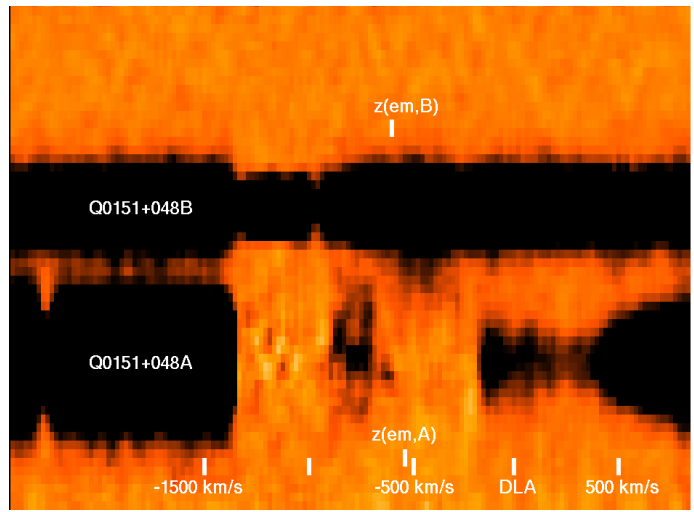} }
	\caption{2D X-shooter spectrum with the slit placed on both QSOs. Short wavelengths are to the left. The centre of the DLA line is marked and chosen as fiducial zero velocity. SPSF subtraction was done for the qA spectrum in the range $-1300$ to $-200$ km s$^{-1}$. There is a clear extended residual seen which we interpret as Ly$\alpha$ emission at $-750$ km s$^{-1}$. The redshifts of both QSOs based on $\ion{H}{\alpha}$ and $\ion{H}{\beta}$ are also marked.}
		\label{twodspec}
\end{figure}

In Fig.~\ref{slitpos} it is seen that
the X-shooter slit follows a line across
the Ly$\alpha$ emitter more closely aligned with it's minor axis. We
would still expect to see some extended emission, but not extending
as far away as for the NOT spectrum. To confirm this expectation we
therefore also performed SPSF subtraction on the X-shooter spectrum.
The result is shown in Fig.~\ref{twodspec},
where we have marked the redshifts
of both QSOs relative to that of the DLA. SPSF subtraction of qA was only performed in the range
$-1300$ to $-200$ km s$^{-1}$ (relative to the DLA redshift).
This procedure left the flux from the host galaxy inside the DLA
trough and clearly shows that the damped line does not at any point
go to zero flux. Also it is seen that the host is precisely
lined up with the position of qA.
The extended Ly$\alpha$ is also here seen to be asymmetric
and to stretch about half way towards qB, consistent with the NB
imaging (Fig.~\ref{slitpos}). The emission is here offset
750 km s$^{-1}$ from the DLA, it has a FWHM of 200 km s$^{-1}$ and
there is no obvious evidence for a rotation curve along this axis.
The difference in velocity offset at the position of the quasar seen
at the two different slit PAs is due to the integration of the velocity
field over different parts of the blob. The velocity measured along
the major axis is less affected by this and we obtain a weighted
combined offset of $640\pm70$ km s$^{-1}$.

There seems to be marginal evidence for a double-emission
profile with a splitting of 140 km s$^{-1}$ but the S/N of the data is
too low to make any conclusive statements about this.
Double-peaked emission line profiles from Ly$\alpha$ were predicted by
\citet{harrington} from a static, un-absorbing, opaque scattering medium.
\citet{verhamme06} obtained the same result using a 3D Ly$\alpha$ radiative transfer code with a Monte-Carlo fitting method, and subsequently used the code to fit Ly$\alpha$ lines in a sample of high redshift Lyman-break galaxies \citep[LBGs;][]{verhamme08}. In this publication, two LBGs had double emission line peaks and were best fitted with a dust-free, nearly static medium. 
Thus, if it can be confirmed that the emission line profile is indeed
double-peaked then this could be indicating that the blob is static rather
than infalling or the signature of an expanding wind.


\begin{table*}
\caption{Voigt-profile fitting (best fits shown in Fig.~\ref{FigFit})
of metal ion transitions in the DLA using \texttt{FITLYMAN}. Ionic
column densities were used to derive individual element abundances
in the neutral gas phase via comparison to the solar neighbourhood
\citep{asplund09}. Multiplets were fitted with same column densities
and turbulent broadening parameter values. The $\ion{Si}{ii}\,\lambda1808$,
$\ion{Fe}{ii}\,\lambda1611$ and $\ion{N}{v}$ lines are not detected
and for those we provide $3\sigma$ column density upper limits (see
text).}
\label{fitting}
\begin{minipage}[t]{\columnwidth}
\centering
\renewcommand{\footnoterule}{}
\begin{tabular}{ c c c c c c c c c c}
\hline\hline
 $\lambda_{obs}$  & $EW_{obs}$ & $\sigma_{EW}$ & $EW_{rest}$ & ID & $z_{abs}$ & log $N$ & $b_{turb}$ & $\rm{[X/H]}$\\
 $\AA$  & $\AA$ & $\AA$ &  $\AA$ &   &  & $\cm^{-2}$ & km s$^{-1}$ & \\
\hline
3634.960 & \mbox{\ldots} & \mbox{\ldots} & \mbox{\ldots} & $\ion{N}{v}$ (1238) & 1.93421 & $<12.71$ & \mbox{\ldots} & \\
3646.650 & \mbox{\ldots} & \mbox{\ldots} & \mbox{\ldots} & $\ion{N}{v}$ (1242) & 1.93421 & $<12.71$ & \mbox{\ldots} & \\
3519.770 & 0.070 & 0.030 & \mbox{\ldots} & $\ion{N}{i}$ (1199) & 1.93424 & $12.95\pm0.12$ & \mbox{\ldots} & \\
\hline
 3820.900  & 0.681 & 0.029 & 0.232 &  $\ion{O}{i}$ (1302) & 1.93426 & $>14.97$  & 20.1$\pm$2.1& $\rm{[O/H]}>-2.06$\\
\hline
    3915.800  & 0.724 & 0.037 & 0.247 & $\ion{C}{ii}$ (1334) & 1.93421 & $>14.46$ & 18.8$\pm$0.2 &  $\rm{[C/H]}>-2.31$\\
    3919.150 & 0.050 & 0.015 & 0.017 & $\ion{C}{ii}^\ast$ (1335.7) & 1.93414 & $12.94\pm0.12$ & \mbox{\ldots} & \\
    4542.600 & 0.928 & 0.040 & 0.316 & $\ion{C}{iv}$ (1548) & 1.93413 & $>14.56$ & 18.9$\pm$0.4 & \\
    4550.155 & 0.813 & 0.035 & 0.277 & $\ion{C}{iv}$ (1550) & 1.93413 & $>14.56$ & 18.9$\pm$0.4 & \\
\hline
    3492.924 & 0.790 & 0.023 & \mbox{\ldots} & $\ion{Si}{ii}$ (1190) & 1.93421 & \mbox{\ldots}\footnote{The $\ion{Si}{ii}\,\lambda 1190$ line is blended with a Ly$\alpha$ forest line and was not included in the overall fit.} & \mbox{\ldots} & \\
    3501.357  & 0.362 & 0.014 & 0.123 & $\ion{Si}{ii}$ (1193) & 1.93421 & $13.92\pm0.03$ & 18.6$\pm$0.4 &\\
    3698.343  & 0.767 & 0.034 & 0.261 & $\ion{Si}{ii}$ (1260) & 1.93421 & $13.92\pm0.03$ & 18.6$\pm$0.4 &\\
    3827.300  &  0.265   & 0.030   & 0.100 & $\ion{Si}{ii}$ (1304) & 1.93421  & $13.92\pm0.03$ & 18.6$\pm$0.4 &\\
    4479.678  & 0.459 & 0.026 & 0.156 & $\ion{Si}{ii}$ (1526) & 1.93421 & $13.92\pm0.03$  & 18.6$\pm$0.4& $\rm{[Si/H]}=-1.93\pm0.04$ \\
    5305.090  & \mbox{\ldots} & \mbox{\ldots} & \mbox{\ldots} & $\ion{Si}{ii}$ (1808) & 1.93421 & $<14.46$ & \mbox{\ldots} & \\
    3540.130 & 0.804 & 0.030 & 0.270   & $\ion{Si}{iii}$ (1206) & 1.93421 & $>13.85$ & 18.8$\pm$0.3 &  \\	
    4089.480 & 0.640 & 0.065 & 0.218 & $ \ion{Si}{iv}$ (1393) & 1.93415 & $>13.75$ & 19.2$\pm$0.7 & \\
    4115.933 & 0.517 & 0.035 & 0.180 & $\ion{Si}{iv}$ (1402) & 1.93415 & $>13.75$ & 19.2$\pm$0.7 & \\
\hline
    3359.492  & 0.103 & 0.024 & 0.035 & $\ion{Fe}{ii}$ (1144) & 1.93421 & $13.60\pm0.02$ & 18.8$\pm$0.3 &  $\rm{[Fe/H]}=-2.24\pm0.03$\\
    4719.540 & 0.144 & 0.035 & 0.049 &  $\ion{Fe}{ii}$ (1608) & 1.93421 & $13.60\pm0.02$  & 18.8$\pm$0.3& \\
    4727.600 & \mbox{\ldots} & \mbox{\ldots} & \mbox{\ldots} &  $\ion{Fe}{ii}$ (1611) & 1.93421 & $<14.67$  &  \mbox{\ldots} & \\
    6967.170 & 0.138 & 0.010 &0.047  &  $\ion{Fe}{ii}$ (2374) & 1.93421 & $13.60\pm0.02$  & 18.8$\pm$0.3& \\
    6991.530 & 0.883 & 0.010 & 0.301 &  $\ion{Fe}{ii}$ (2382) & 1.93421 & $13.60\pm0.02$  & 18.8$\pm$0.3& \\
    7589.770  & 0.410 & 0.013 & 0.140 &  $\ion{Fe}{ii}$ (2586) & 1.93421 & $13.60\pm0.02$  & 18.8$\pm$0.3& \\
\hline
    4902.464  & 0.321 & 0.041 & 0.110 &  $\ion{Al}{ii}$ (1670) & 1.93423 & $12.46\pm0.02$  & 18.8$\pm$0.2&  $\rm{[Al/H]}=-2.33\pm0.03$\\
    5442.160	& 0.09 & 0.023 & 0.030 & $\ion{Al}{iii}$ (1854) & 1.93423 & $12.15\pm0.07$ & $18.8\pm0.2$ & \\
    5465.850	& 0.060 & 0.030  & 0.020	& $\ion{Al}{iii}$ (1862) & 1.93423 & $12.15\pm0.07$ & $18.8\pm0.2$ & \\
\hline
     8205.080  & 1.611 & 0.010 & 0.550 &  $\ion{Mg}{ii}$ (2796) & 1.93421 & $>13.79$  & 18.8$\pm$0.3&  $\rm{[Mg/H]}>-2.15$\\
     8226.140  & 1.380 & 0.011 & 0.470 &  $\ion{Mg}{ii}$ (2803) & 1.93421 & $>13.79$ & 18.8$\pm$0.3& \\
     8371.210 & 0.058 & 0.021 & 0.020 & $\ion{Mg}{i}$ (2852) &	1.93421  & $11.19\pm0.11$ & \mbox{\ldots} & \\
\hline
\end{tabular}
\end{minipage}
\end{table*}

\section{Absorption line analysis}

The X-shooter absorption line spectra are plotted in Fig.~\ref{Figuvb}.
The medium resolution of X-shooter allows us to perform absorption line
profile fitting and determine column densities from lines that
are not heavily saturated. Our second main driver for the
X-shooter spectroscopy was therefore to derive ionic column densities and element abundances for the DLA system observed in the spectrum of qA.

\begin{figure*}
  \centering
{\includegraphics{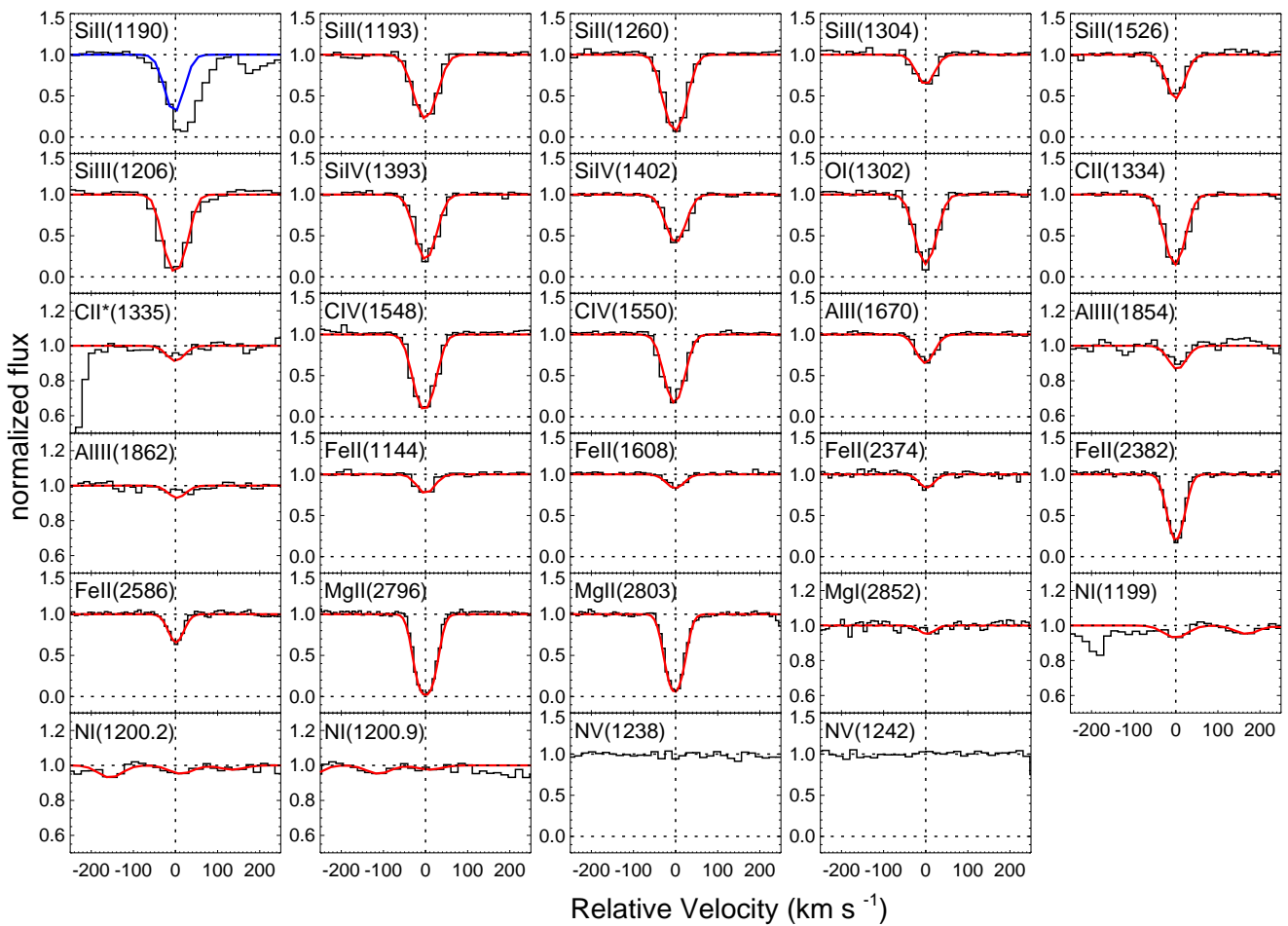}}
\caption{Voigt-profile fits (red overlay) to the metal absorption
  lines of the DLA in the spectrum of qA. The derived parameters are
  provided in Table~\ref{fitting}. The vertical dashed black line is
  our adopted zero velocity corresponding to $z=1.93421$. For
  $\ion{Si}{ii}\,\lambda1190$, we did not fit the line but show the
  expected line in blue. For $\ion{N}{i}$, we only list the strongest
  line in the table but the fit was performed simultaneous to the
  triplet as shown.}
         \label{FigFit}
  \end{figure*}

Element abundances are important for our understanding
of this system, first because it has been suggested that the DLA
metallicity is related to the mass and luminosity of
the DLA galaxy \citep{moller2004,ledoux, fynbo2008, prochaska2008}
second because it can provide us with limits on the QSO/DLA distance
(\S \ref{ionizationsection}) which are needed in order to determine the
arrangement of the QSO/DLA/Ly$\alpha$ emitter system
in 3D space rather than redshift space.
In addition several systems at lower redshifts are
also seen in the X-shooter spectra and for completeness
we also identify those (\S \ref{absorbersection}).

\subsection{Absorption lines of the DLA system}

\citet{moller} reported an ${\rm H}^0$ column density of the DLA of
log $N_{{\rm H}^0}=20.36$. In our higher quality data we determine
a column of log $N_{{\rm H}^0}=20.34\pm0.02$ $\cm^{-2}$ at $z=1.93421\pm 0.00001$. The Voigt profile fit to the DLA is overplotted in blue on the normalized spectrum in Fig.~\ref{Figuvb}. At the same
redshift we also identify several metal lines in the spectrum of qA,
of which none are seen in the spectrum of qB down to a 3$\sigma$ upper limit of 0.08\,\AA{} (see Fig. \ref{Figuvb}). We identify lines from ${\rm O}^0$, ${\rm C}^+$, ${\rm C}^{+\ast}$
${\rm C}^{3+}$, ${\rm Si}^+$, ${\rm Si}^{2+}$, ${\rm Si}^{3+}$,
${\rm N}^0$, ${\rm Fe}^+$, ${\rm Al}^+$, ${\rm Al}^{2+}$,
${\rm Mg}^0$ and ${\rm Mg}^+$.
Equivalent widths and corresponding 1 sigma errors of
those lines are listed in Table~\ref{fitting}.

We used the Voigt profile fitting \texttt{FITLYMAN} package in MIDAS
\citep{fontana} and adopted the list of atomic data, laboratory
wavelengths and oscillator strengths from \citet{morton2003}.
\texttt{FITLYMAN} finds the
best global fit using a $\chi^2$ minimization algorithm
which includes the spectral resolution and
returns best fit parameters for redshift, column
density and Doppler turbulent broadening as well as errors
on each quantity. It may be used to fit single lines one
at a time, or one may (optionally) choose to fit several
lines with the same parameters (one may e.g. insist that
all the chosen lines have the same redshift or
turbulent broadening).

We employed \texttt{FITLYMAN} in two ways. First we chose to fit all
lines with the same redshift and turbulent broadening. The resulting
fits were mostly very good, but for a few atoms and ions (e.g. ${\rm O}^0$
and ${\rm Si}^{3+}$) the fits were slightly offset from the data. We
therefore chose to fit each ionic species independently allowing them
to have different redshifts and turbulent broadening. The resulting
column densities did not change significantly from one procedure to
the other so our results are not depending on this choice. The
\texttt{FITLYMAN} output parameters are listed in Table~\ref{fitting}
and the resulting fits (red) are shown overlaid on the data in
Fig.~\ref{FigFit}. Note that the $\ion{Si}{ii}\,\lambda 1190$
line is in the Lyman forest and clearly blended with an
intervening line. That line was not used for the fit, but
the predicted line is plotted for completeness.

While the signal-to-noise ratio of our X-shooter spectrum is
  excellent, its resolving power is not high enough to resolve the
  lines which at this resolution may have hidden velocity
  sub-structure. Accurate column densities can only be derived from
  weak lines while the column densities of stronger lines may be
  underestimated due to hidden saturation. For this reason, we
  consider our fit results for $\ion{O}{i}$, $\ion{C}{ii}$ and
  $\ion{Mg}{ii}$ to be lower limits. For $\ion{Si}{ii}$,
  $\ion{Al}{ii}$, $\ion{Fe}{ii}$, $\ion{C}{ii}^\ast$, $\ion{Mg}{i}$
  and $\ion{N}{i}$, hidden saturation could be an issue in case
  additional components with $b$ values narrower than 3 km s$^{-1}$
  are present. The non-detection of $\ion{Si}{ii}\lambda 1808$ sets an
  upper limit of $10^{14}$ on the column density of any component with
  a $b$ of 3 km s$^{-1}$. \citet{ellison} presented a spectrum of qA
  of high resolution ($\approx 6$ km s$^{-1}$ FWHM) with similar
  detection limits as the X-shooter spectrum. By way of comparison, we
  note that our column densities of ${\rm Si}^+$, ${\rm Al}^+$, ${\rm
    Fe}^+$, ${\rm N}^0$ and ${\rm C}^{+\ast}$ are 0.09, 0.11, 0.10,
  0.11 and 0.06 dex lower than those measured by \citet{ellison} which
  we take as the error induced by the uncertainty on our derived $b$
  value. This shows that narrow hidden components account for less than 0.1 dex of
the total column densities of these ions.

For non-detected lines, we report in Table~\ref{fitting} $3\sigma$
column density upper limits. For those, we computed the column
densities assuming them to be on the linear part of the
curve-of-growth rather than from the Voigt-profile fitting.

\begin{figure}
\centering
{\includegraphics[angle=-90,width=\columnwidth,bb=50 60 580 630,clip=]{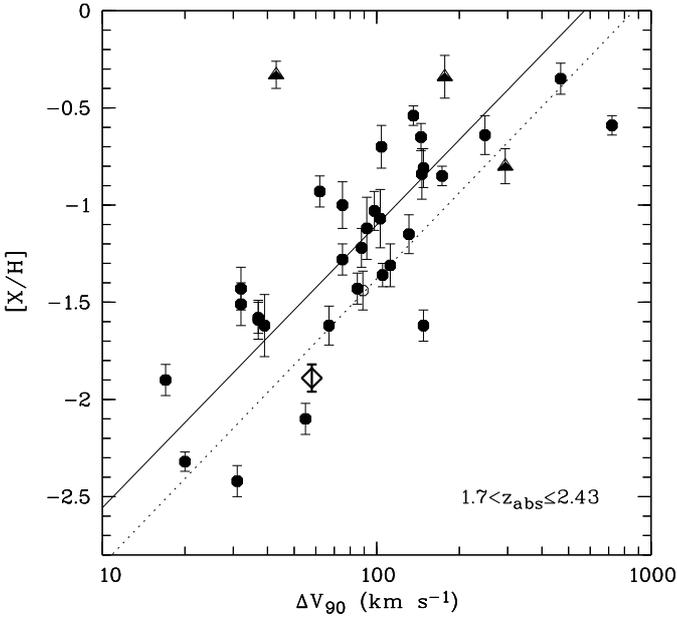} }
\caption{
Average metallicity vs. velocity width of 35 DLA low ionization line profiles from \citet{ledoux}. The solid line represents the linear least square bisector fit. The
DLA from the present work is marked by the large diamond. For details
on this figure see \citet{ledoux}.}
\label{ledouxvel}
\end{figure}

In the last column of Table~\ref{fitting}, we adopt the solar
abundances from \citet{asplund09} to derive metallicities relative to
solar. We find a silicon abundance, which is commonly used to define
DLA metallicities, of $\rm{[Si/H]}=-1.93\pm 0.04$, suggesting a
metal-poor system (see Fig.~\ref{ledouxvel}, \citealt{ledoux}; \citealt{prochaska}). We determine a line velocity width of 64 km
s$^{-1}$ as defined in \citet{ledoux} using $\ion{Fe}{ii}\,\lambda
2382$ which, after correction for the instrumental resolution of 27 km
s$^{-1}$, leads to 58 km s$^{-1}$. In Fig.~\ref{ledouxvel}, we plot
this together with the 35 DLAs in the redshift range 1.7 to 2.43 taken
from \citet{ledoux}.

One can see that the point follows the general metallicity/velocity
width relation at $z=2$, but that it falls within the 10--15\% lowest
metallicities, and that it lies on the high $\Delta$V side of the
bisector fit (in the widest 15\%).

Collisions with electrons are the dominant excitation mechanism to
populate the upper fine-structure energy level of the ${\rm C}^+$
ground state \citep[see also][]{williams}. The
$\ion{C}{ii}^\ast\,\lambda1335.7$ line can therefore be used to set an
upper limit on the particle density $n_{{\rm H}^0}$. Unfortunately,
the $\ion{C}{ii}$ line only provides a column density lower limit due
to saturation, but assuming solar abundance ratio of Si/C we can use
${\rm Si}^+$ as a proxy and compute the ${\rm C}^+$ column density. We
find that: log $N_{{\rm C}^{+\ast}}/N_{{\rm C}^+}=-1.9$, and from fig.
4 of \citep{silva}, we determine a particle density of $n_{{\rm
    H}^0}=25$ cm$^{-3}$.


\subsection{Photo-ionization model of the DLA}
\label{ionizationsection}

Based on the column densities reported above, we now make a simple
photo-ionization model of the absorbing cloud. The main purpose of the
model is to constrain the distance between qA and the DLA. For the
model, we use Cloudy version 07.02 described by \citet{ferland}. The
Cloudy calculations are based on the following assumptions. The cloud
is assumed to be a plane-parallel slab of gas with total neutral
hydrogen column density log $N_{{\rm H}^0}=20.34$, metallicity 1/100th
of solar, solar relative abundance ratios, and a particle density in
the neutral phase ($n_{{\rm H}^0}$) of 25 cm$^{-3}$. The radiation
field is composed of the Cosmic Microwave Background at $z=1.934$ and,
towards one side of the cloud, the radiation spectrum of a typical AGN
with specific luminosity at the Lyman limit $L_{\nu _0}=30.9$ erg
s$^{-1}$ Hz$^{-1}$. The latter value has been determined using the
flux-calibrated spectrum obtained from NOT in August 2008 and a
best-matched spectral slope of $-1.76$ between rest-frame 850 and
1600\,\AA\ (see \citealt{telfer}).

Using this model, we performed a series of photo-ionization
  equilibrium calculations for a range of QSO-DLA distances and
  computed ionic column densities.
  Keeping in mind that this is a simple model we can nevertheless
  apply our observational constraints to get an estimate of the
  distance.

From log $N_{{\rm Al}^{2+}}/N_{{\rm Al}^+}=0.31 \pm 0.09$, we
  find agreement for $30\pm 11$ kpc while the upper limit log $N_{{\rm
      N}^{4+}}/N_{{\rm N}^0}<0.14$ sets a hard lower limit on the
  distance of $28$ kpc. This model is therefore in best agreement with
  a distance of 30-40 kpc assuming no local source of ionization.  The
  column densities of the higher ionization gas (${\rm Si}^{2+}$,
  ${\rm Si}^{3+}$ and ${\rm C}^{3+}$) are all lower limits but we note
  in passing that those limits are all consistent with the above
  distance estimate.



\subsection{Intervening absorbers}
\label{absorbersection}

The primary purpose of the X-shooter spectroscopy was to characterize
the DLA and the QSO. In addition to the DLA four other $\ion{C}{iv}$
absorption systems at redshifts 1.2939, 1.3360, 1.5613, and 1.6192
were identified in the spectra. All four $\ion{C}{iv}$ doublets are seen in
both spectra (as seen in Fig.~\ref{Figuvb}) indicating that $\ion{C}{iv}$ absorbers at
those redshifts typically have sizes larger than 30 kpc and covering
factors close to unity. This is consistent with previous studies
\citep{rauch,ellison2004}. One of the absorbers ($z=1.6192$) is also seen as
$\ion{Si}{iv}$ absorber in the sightline towards qA, but not towards qB,
which might indicate that $\ion{Si}{iv}$ absorbers extend to smaller
distances than $\ion{C}{iv}$ absorbers.

It is striking in Fig.~\ref{Figuvb} to see the one-to-one
correspondence between the lines in the Lyman forest of the two
QSOs \citep[see][]{smette,smette1995}. The few lines which are not reproduced in the qB spectrum all 
belong to the DLA system.

\section{
What is the Q\,0151+048A{\&}B/DLA/Ly$\alpha$-blob system?}

This complex system has been the target of several studies because it
brings together several types of objects which at this redshift are
usually seen only separately. The mix of objects has made
the interpretation
difficult, yet at the same time the interaction between the objects
allow us to extract information which is not normally accessible and
can only be obtained for this unique system.
Earlier interpretations were hampered by the lack of certain vital
information but the new X-shooter data have now allowed us a more
complete view of this system.

\subsection{Geometry of the system}

Fig.~\ref{twodspec} summarizes what we know about the redshifts
of the components. The issue with the redshifts of the QSOs has been
resolved causing qA to be moved 740 km s$^{-1}$ towards higher redshift.
This means that qA is now offset only 100 km s$^{-1}$ from the
Ly$\alpha$ emission (at the position of qA) in the NOT spectrum, and
200 km s$^{-1}$ in the X-shooter spectrum. The difference is
presumably arising from the different way the velocity field of
the emitter has been averaged over the slit in the two
observations. The velocity offset between the DLA and the
Ly$\alpha$ blob is much larger ($640 \pm 70$ km s$^{-1}$).
All taken together the logical conclusion is that the blob is
associated with qA and its host galaxy rather than with the DLA as it
was previously thought. The end-to-end size of the blob along it's
major axis is about 5 arcseconds corresponding to 42 kpc, and 28 kpc
along its minor axis while the
projected distance from qA to qB is 27.7 kpc.
Based on the ionization state of the gas we find that the DLA
cloud is placed along our line of sight towards qA at a
distance larger than $\sim30$ kpc.

\subsection{Dynamical state of the system}
\label{dynamicalsection}

  \begin{table}
\caption{BH, total masses and edingtton ratios of the QSOs. Masses are provided in logarithmic scale.} 
\label{masses} 
\centering     
\begin{tabular}{@{}l c c c @{}}  
\hline\hline                        
	&	&	qA	& qB \\
	\hline
FWHM H$\beta$ (km s$^{-1}$) & & $3577\pm150$ & \mbox{\ldots}\\
FWHM $\ion{Mg}{ii}$ (km s$^{-1}$) & & $2790\pm110$ & $1840\pm130$ \\
FWHM $\ion{C}{iv}$ (km s$^{-1}$) & & $3875\pm130$ & $2480\pm110$ \\
$\lambda L_{\lambda}$ ($10^{44}$ erg s$^{-1}$)& (5100\,\AA)  & $170$ & $5.8$ \\
$\lambda L_{\lambda}$ ($10^{44}$ erg s$^{-1}$) & (3000\,\AA)& $2100$ & $105$ \\
$\lambda L_{\lambda}$($10^{44}$ erg s$^{-1}$) & (1450\,\AA) & $1400$ & $67$ \\
$M_{BH}$	/$M_\odot$  & (H$\beta$)& 	$9.12$ &	\mbox{\ldots} \\
$M_{BH}$	/$M_\odot$  & ($\ion{Mg}{ii}$) & $9.37$ & $8.38$ \\
$M_{BH}$	/$M_\odot$ &($\ion{C}{iv}$) & $9.50$ & $8.36$ \\
$M_{tot}$/$M_\odot$ & & $13.74$ & $13.13$ \\
log $L_{\rm{bol}}/L_{\rm{Edd}}$ & 	& $-0.24$ & $-0.54$ \\
\hline
\end{tabular}
\end{table}

The systemic redshift difference between qA and qB is very small
(60 km s$^{-1}$ and consistent with zero to within less than 2$\sigma$)
suggesting that their relative movement is in the plane of the sky.
This would imply that their physical separation is the same as their
projected separation.
The Ly$\alpha$ blob has a velocity gradient along its major axis and
through qA. This may be caused by infalling gas which is being ionized
where it falls into the ionizing cones of qA as it was described
in the models by \citet{weidinger04,weidinger} but it could also
be the signature of a rotating disk.
From Fig.~\ref{velocity}, and
considering the resolution of the data along the slit corresponding
to a smoothing length of 13 kpc, we see that the current data are
indeed consistent with the rotation curve of a large spiral
galaxy in todays universe. A spectrum with better spatial resolution,
and along the major axis of the blob, would be needed to
discriminate between the two possibilities.

The DLA absorber is a separate object which is moving towards
the Ly$\alpha$ blob with a relative velocity of $640 \pm 70$ km s$^{-1}$.

\subsection{Black Hole masses, DM Halo masses and Eddington ratios}

\begin{figure}
  \centering
{\includegraphics[width=\columnwidth,clip=]{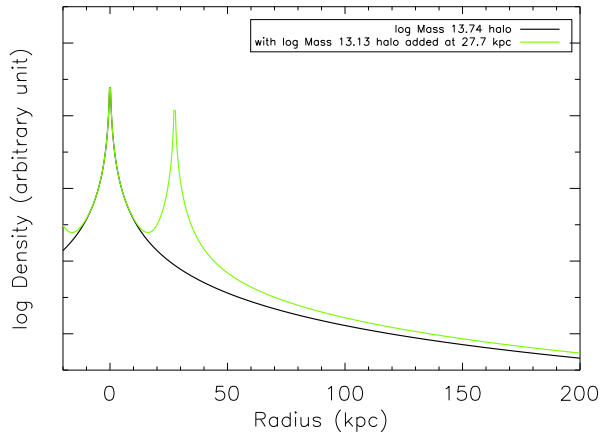}}
     \caption{Density profile of the two DM halos around qA and qB
assuming that they have standard NFW profiles.}
         \label{NFW}
  \end{figure}

The mass of the DM halo of the QSO Q1205-30 at $z=3.04$ was determined
from a model of the gas falling into its potential well and was
found to be in the range log $M_{tot}/M_\odot = 12.3-12.8$
\citep{weidinger04}. As discussed in \S \ref{dynamicalsection} it is not clear
in our case if the dynamical signature of the gas is caused by
infall or rotation so we shall use a different method to determine
the DM halo mass.

First we derive Black hole (BH) masses following \citet[formulae 5 \& 7]{vestergaard06} and use the relation between BH mass, H$\beta$ and $\ion{C}{iv}$ emission lines width and the continuum luminosity at 5100\,\AA \ and 1450\,\AA \ rest wavelength. We also estimated BH mass using the relation between $\ion{Mg}{ii}$ and continuum luminosity at rest-frame 3000\,\AA \ \citep[formula 1]{vestergaard}. For the line widths we use the FWHM of the fitted gaussian for each emission line and we list those in Table \ref{masses}. We corrected the FWHM of each line for spectral resolution as described in \citet{peterson}. For qB we do not have the H$\beta$ emission line width, therefore, we have no mass estimation from that line.

To obtain the continuum luminosity at 5100, 3000 and 1450\,\AA \ we
use the $H$-band magnitudes of the two QSOs from \citet{warren} and the $I$ and $B$-band magnitudes from \citet{fynbo2}. We also checked for line contamination in these band and for this we used the composite spectrum from \citet{vanden}. We subtracted emission line components from the spectrum and found the contamination to be 14, 15 and 7\% in the $H$, $I$ and $B$ bands respectively. After subtracting the contamination the
resulting BH masses are listed in Table \ref{masses}.
It has recently been established that BH mass
correlates well with the underlying total gravitational mass
(presumably dominated by the dark matter) of the host halo
\citep{bandara}. We use this correlation to finally obtain the
corresponding halo mass of each QSO by using their mean BH mass and those are also listed
in Table \ref{masses}.

In Fig.~\ref{NFW} we plot the density profile of the two halos assuming
that they have NFW profiles \citep{navarro} and that their projected
distance (27.7 kpc) is also their real proper distance as argued in
\S \ref{dynamicalsection}. It is seen that qB is inside the halo of qA,
but that it is still clearly defined as a separate sub-halo.
We used the prescription from \citet{nemmen} and 
\citet{gavignaud} to determine bolometric luminosities and
eddington ratios (provided in Table \ref{masses}) respectively.

\section{Discussion and Conclusions}
 
In this paper we have presented spectra of the two QSOs Q\,0151+048A
and Q\,0151+048B as well as of a Ly$\alpha$ blob at the same position.
We obtained spectra of the blob at two different position angles.
QSO redshifts determined from H$\beta$ and H$\alpha$ change the
previously reported redshifts by 740 km s$^{-1}$ in addition to
the much smaller corrections previously suggested by \citet{tytler}.
Our first conclusion is that the corrections to QSO redshifts based
on rest-frame UV lines can be much larger than previously thought and
that other QSO absorption systems of this kind (i.e. with large
inverted redshift offsets) presumably will turn out to find the same
resolution.

The new redshift of qA means that the Ly$\alpha$ blob is now more
likely belonging to this QSO. We determine a Black Hole mass of
qA of $10^{9.33}~M_\odot$ and a corresponding DM halo mass of
$10^{13.74}~M_\odot$. The Ly$\alpha$ blob is sitting in the
potential well
of this halo. The blob has a size of 42 by 28 kpc, it has a clear
signature of a strong velocity gradient along its major axis which
is consistent with the rotation curve of a large disk galaxy in
todays universe but may alternatively be caused by in- or out-flow.
In the X-shooter spectrum along the minor axis which has higher
spatial resolution there is no velocity gradient of the blob,
but there is a hint that the Ly$\alpha$ emission may be double
peaked.

Interpolating in the merger tree simulations of
\citet{parkinson} we find that the halo of qA in the mean will end up
in a halo with a mass of $6\times10^{14}$ M$_\odot$ at $z=0$. We
report bolometric accretion rates of both quasars and find that they
accrete at 60\% and 30\% of their Eddington limits. This places
both qA and qB consistently on the relation shown in fig.~4
of \citet{gavignaud}.

From the aspect of the two quasars they therefore appear to be very
typical for quasars at that redshift, and their environment is
apparently a proto-typical pre-cursor of what will become something very
close to our local super-cluster at $z=0$. What makes this system special
is the accidental alignment of many different types of objects so
close together. The physical interactions between
the dual quasar, the Ly$\alpha$ blob, the host galaxies and the PDLA
absorber makes this a unique laboratory to study what seems to be
what our local super-cluster looked like at $z=2$.

The metallicity of the DLA inferred from the X-shooter medium
resolution spectrum is $\rm{[Si/H]}=-1.93\pm0.04$, in the lowest 15\%
of DLAs at this redshift. This stands in contrast to the recent
result by \citet{ellison} who reported that PDLAs in general have
higher metallicity than intervening DLAs. 

Based on the ionization state of the gas we model the
physical state of the DLA and find a best fit at a
distance of 30-40 kpc from qA assuming that contribution
of local star-formation can be ignored. Under the same
assumption, and using a different method,
\citet{ellison} found that the distance must be larger than
160.5 kpc. In case there is local star formation close to the DLA
our model would only provide a lower limit and would be
consistent with the result of \citet{ellison}.

The DLA falls nicely on the relation of velocity-width
vs metallicity \citet{ledoux}. Using the $\rm{[X/H]}$ vs the $R$-band magnitude 
relation derived in that paper we predict that the DLA should have an
R-band brightness of about $m_R=30$ and a Ly$\alpha$ flux of around
$2\times10^{-18}$ erg s$^{-1}$ cm$^{-2}$. This object is therefore
not observable in emission in our data, consistent with the
interpretation of the Ly$\alpha$ blob belonging to the host galaxy and not to the DLA.

\citet{heckman} reported the detection of ``fuzz''
around radio loud quasars and argued that the fuzz had to be clumpy
on small scales with a very small filling factor. \citet{weidinger04} 
reported similar fuzz around a radio quiet quasar and showed that
its velocity profile and morphology implied that it was in a state
of infall. That quasar also exhibit a strong absorber close to its
emission redshift, a system which in current terminology may be named
a proximate sub-DLA. The clumpiness on small scales and the small filling
factor predicted by \citet{heckman} matches the few known PDLA systems
well (qA, \citealt{leibundgut}, \citealt{weidinger04},
\citealt{hennawi}) and like \citet{hennawi} we consider it likely
that PDLA Ly$\alpha$ blobs in some ways are related to the fuzz around
radio loud quasars. However, there are significant differences in
terms of energy output, in the measured Ly$\alpha$ line widths (which
is much narrower for qA), in the mechanical interaction with the
radio jet and in the fact that metal emission lines are detected in
the fuzz around radio loud object. The case of qA is
therefore more similar to that of \citet{weidinger04} and the negative
velocity of the DLA is an indication that it is in a similar state
of infall.

We detect emission in the trough of the DLA which we interpret
as the signature of the quasar host galaxy. It is interesting to
note that the hosts were not detected in any of the three above
mentioned cases which are all at $z=3$. This is consistent with recent
reports that there is a rapid transition of Ly$\alpha$ selected
objects observed at a redshift of about 2.5 \citep{nilsson09,nilssonm}.

We briefly report the detection of four intervening $\ion{C}{iv}$ systems
seen in both sightlines of which one system has $\ion{Si}{iv}$ absorption seen
only in a single sighline. Also we see rich Lyman forests with
near identical absorption signatures. This QSO pair is therefore
well suited for a multi-sightline study of both Lyman forest
absorbers and of intervening metal absorbers.
 
\begin{figure*}
   \centering
   {\includegraphics[angle=180]{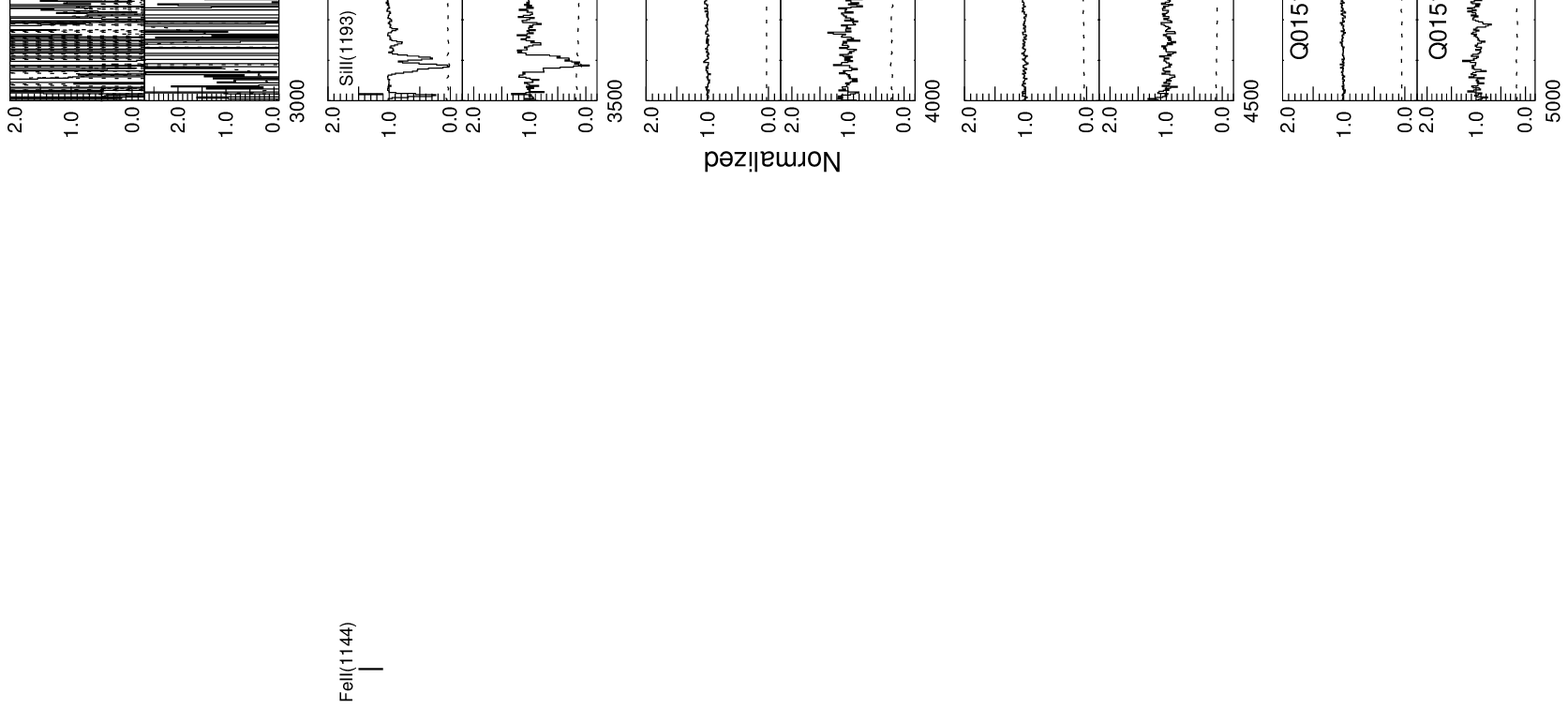}}
      \caption{X-shooter UVB-arm spectra from $3000-5500\AA$ of qA in top panel and qB in bottom panel in each grid. Solid black lines with corresponding IDs in Q\,0151+048A spectrum are from the DLA. The dashed blue line in Q\,0151+048 spectrum
shows the Voigt profile fit to the DLA with an inferred column
density of log $N_{{\rm H}^0}=20.34\pm0.02$ for a redshift of 1.93421. Dashed red lines in both spectra are from the intervening absorbers with their corresponding IDs. The dotted line represents 1$\sigma$ error on each spectra.}
          \label{Figuvb}
   \end{figure*}


\begin{acknowledgements}
The Dark Cosmology Centre is funded by the Danish National Research Foundation. We are thankful to the VLT X-shooter team. X-shooter is funded with contributions from ESO, Danish, Dutch, Italian and French partners. We are thankful to the anonymous referee for the constructive comments. We are thankful to Sara Ellison, Marianne Vestergaard and Alain Smette for helpful comments.

\end{acknowledgements}

\bibliographystyle{aa}
\bibliography{Qso-draft8.bib}{}

\end{document}